\documentclass[aps,pra,preprint,superscriptaddress,longbibliography]{revtex4-2}

\usepackage{amsmath}
\usepackage{amsthm}
\usepackage{amsfonts}
\usepackage{amssymb}
\usepackage{xcolor,graphicx}
\usepackage{tikz}
\usepackage{color}
\usepackage[pdftex]{hyperref}
\usepackage{longtable}
\usepackage{array}
\usepackage{dsfont}
\usepackage{bm}

\begin{document}

    \title{Conditional Gaussian filtering by Arthurs--Kelly readout in a three-mode cluster wire}

    \author{B.~M. Rodr\'iguez-Lara}
    \email[e-mail: ]{blas.rodriguez@gmail.com}
    \affiliation{Universidad Polit\'ecnica Metropolitana de Hidalgo, Tolcayuca, Hidalgo 43860, Mexico.}

    \author{J.~A. Mendoza-Fierro}
    \email[e-mail: ]{jmendoza@tec.mx}
    \affiliation{Tecnologico de Monterrey, Escuela de Ingenier\'ia y Ciencias, Bulevar Tomás Fernández 8945, 32470 Ciudad Juárez, Chihuahua, Mexico.}

    \date{\today}

    \begin{abstract}
    Measurement readout programs the Gaussian operation implemented by a continuous-variable cluster wire.
    For the standard three-node wire, momentum homodyne readout gives an additive parity channel with finite-squeezing noise.
    We classify the pattern-resolved input--output covariance transformations generated by calibrated uncorrelated Arthurs--Kelly readout on one or both consumed nodes.
    Any finite Arthurs--Kelly position record produces a conditional Gaussian filter whose gain and residual covariance depend on the input covariance entries, rather than an input-independent additive Gaussian channel.
    Writing $\mathrm{A}$ for Arthurs--Kelly readout and $\mathrm{H}$ for homodyne readout, the readout pattern selects the filtered sector, with $\mathrm{AH}$ selecting position, $\mathrm{HA}$ selecting momentum, and $\mathrm{AA}$ retaining both.
    The hierarchy $0 < \tau_{\mathrm{AA}} < \min \left\{ \tau_{\mathrm{AH}}, \tau_{\mathrm{HA}} \right\} < 1 = \tau_{\mathrm{HH}}$ shows that finite Arthurs--Kelly position records contract the gain-transferred covariance area.
    Our results identify calibrated uncorrelated Arthurs--Kelly readout as a measurement-level method for covariance-sensitive Gaussian filtering inside a fixed cluster graph.
    \end{abstract}
    
\maketitle
\newpage
    

\section{Introduction}
\label{sec:Sec1}

Continuous-variable measurement-based architectures implement Gaussian operations through a resource graph, local Gaussian readout, and outcome-dependent displacement corrections~\cite{Menicucci2006p110501,Zhang2006p032318,vanLoock2007p032321,Gu2009p062318,Menicucci2011p042335}.
The readout layer actively shapes the implemented operation~\cite{Alexander2016p062326,Booth2023p1146}.
Changing the measured quadratures can change the Gaussian transformation on the retained modes even when the graph and resource covariance remain fixed~\cite{Gu2009p062318,Alexander2014p062324,Alexander2016p062326}.
Gaussian postselection can emulate noiseless amplification or attenuation in continuous-variable quantum key distribution~\cite{Fiurasek2012p060302,Walk2013p020303}, while teleportation with Gaussian postselection can be written as an effective noiseless operation followed by a modified Gaussian channel~\cite{Blandino2016p012326}.
Measurement-based noiseless amplification similarly requires an effective-transformation characterization of the operation selected by the measurement record~\cite{Zhao2017p012319}.

The three-node line graph~\cite{Yokoyama2015p032304} gives an elementary input--output covariance transformation of a one-dimensional continuous-variable cluster wire, connecting one injected Gaussian input to one retained output through two consumed readout nodes~\cite{Gu2009p062318,Alexander2014p062324}.
Momentum homodyne readout on both consumed nodes gives the reference additive parity channel with finite-squeezing noise~\cite{Gu2009p062318,Alexander2014p062324,Ukai2011p240504}.
Replacing homodyne readout on one or both consumed nodes by finite joint-quadrature readout keeps additional measurement information in the conditioning record~\cite{Arthurs1965p725,Appleby1998p1491,Gampel2023p012420}.
That record enters the Schur complement and can make the retained covariance depend on the input covariance, producing a conditional Gaussian filter rather than an input-independent additive Gaussian channel.

Arthurs--Kelly readout provides a controlled replacement for the consumed homodyne nodes of the three-node cluster wire.
It is a calibrated joint-quadrature Gaussian measurement with explicit position and momentum readout variances that contains momentum homodyne readout as the singular limit where the position record becomes uninformative while the momentum record becomes sharp~\cite{Arthurs1965p725,Appleby1998p1491,Arvind2020p126543}.
A node-resolved Arthurs--Kelly description of the two consumed readouts contains the homodyne additive channel as a limiting pattern and allows finite-readout patterns that produce conditional Gaussian filters.

Here we classify the four readout patterns generated by node-resolved Arthurs--Kelly readout.
We introduce the cluster-wire covariance, the calibrated uncorrelated Arthurs--Kelly readout model, and the Gaussian conditioning rule in Sec.~\ref{sec:Sec2}.
In Sec.~\ref{sec:Sec3}, we derive the pattern-resolved transformations and identify the finite-readout filters.
The singular homodyne limit gives the $\mathrm{HH}$ reference pattern, an additive Gaussian channel that implements parity gain with additive finite-squeezing noise.
Any finite Arthurs--Kelly position record changes the operation class from an input-independent additive Gaussian channel to a covariance-dependent conditional Gaussian filter.
The readout pattern selects the filtered covariance sector, with $\mathrm{AH}$ selecting position, $\mathrm{HA}$ selecting momentum, and $\mathrm{AA}$ retaining both sectors.
In Sec.~\ref{sec:Sec4}, we use output-area and gain-area diagnostics to quantify this filtering structure.
The gain hierarchy $0 < \tau_{\mathrm{AA}} < \min \left\{ \tau_{\mathrm{AH}}, \tau_{\mathrm{HA}} \right\} < 1 = \tau_{\mathrm{HH}}$ shows that finite Arthurs--Kelly position records contract the gain-transferred covariance area.
This identifies calibrated uncorrelated Arthurs--Kelly readout as a measurement-level method for covariance-sensitive Gaussian filtering inside a fixed cluster graph.
We close with a summary in Sec.~\ref{sec:Sec5}.

\section{Gaussian cluster wire with Arthurs--Kelly readout}
\label{sec:Sec2}

\begin{figure}[t]
    \centering
    \includegraphics[width= 0.75 \textwidth]{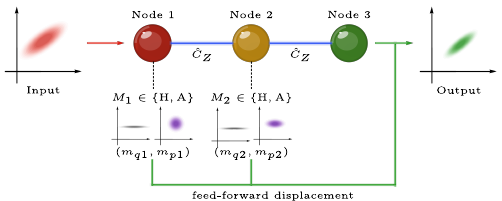}
    \caption{
    Three-mode continuous-variable cluster wire with configurable readout.
    The injected input enters node $1$, nodes $1$ and $2$ are consumed by readout, and node $3$ carries the retained output.
    The blue links denote the controlled-phase network $\hat{C}_{Z}(\bm{V}_{\mathrm{L}})$.
    Each consumed node has readout choice $M_j\in\{\mathrm{H},\mathrm{A}\}$ with outcome classical records $(m_{qj},m_{pj})$.
    The green path is the feed-forward displacement applied to the retained output mode.
}
    \label{fig:Fig1}
\end{figure}

We consider the minimal continuous-variable cluster wire that teleports an arbitrary one-mode Gaussian input through a three-node line graph.
The input is injected into node $1$, nodes $1$ and $2$ are consumed by momentum homodyne measurements, and node $3$ carries the output after feed-forward~\cite{Menicucci2006p110501,Gu2009p062318,Alexander2014p062324,Ukai2011p240504}.
For finite cluster squeezing, the reduced wire description implements the parity channel up to additive Gaussian noise set by the finite nullifier variances~\cite{Gu2009p062318,Alexander2014p062324,Ukai2011p240504}.
We replace one or both consumed homodyne readouts by calibrated Arthurs--Kelly (AK) joint-quadrature readouts~\cite{Arthurs1965p725,Arvind2020p126543}, as shown in Fig.~\ref{fig:Fig1}, and determine the resulting Gaussian input--output map.

We label the readout pattern by
\begin{align}
    \mathcal{R} \in \left\{ \mathrm{HH}, \mathrm{AH}, \mathrm{HA}, \mathrm{AA} \right\},
\end{align}
where $\mathrm{H}$ and $\mathrm{A}$ denote momentum homodyne readout and calibrated AK readout on a consumed node.
For each pattern, we test whether the retained mode defines an input-independent additive Gaussian channel~\cite{Holevo2007p1,Weedbrook2012p621},
\begin{align}
    \bm{\sigma}_{\mathrm{out}}^{(\mathcal{R})} =&~ \bm{G}_{\mathcal{R}} \bm{\sigma}_{\mathrm{in}} \bm{G}_{\mathcal{R}}^{\mathrm{T}} + \bm{N}_{\mathcal{R}},
\end{align}
with $\bm{G}_{\mathcal{R}}$ and $\bm{N}_{\mathcal{R}}$ independent of $\bm{\sigma}_{\mathrm{in}}$.
If finite-record conditioning makes the gain $\bm{G}_{\mathcal{R}}\left( \bm{\sigma}_{\mathrm{in}} \right)$ and residual covariance $\bm{N}_{\mathcal{R}}^{(\mathrm{cond})}\left( \bm{\sigma}_{\mathrm{in}} \right)$ depend on the input covariance matrix, the retained operation defines a conditional Gaussian filter~\cite{Genoni2016p331}.

We use the standard covariance-matrix description of Gaussian states~\cite{Arvind1995p471,Weedbrook2012p621}.
We collect canonical quadratures in the three-mode phase-space operator vector,
\begin{align}
    \begin{aligned}
        \hat{\bm{x}} =&~ \left( \hat{q}_{1}, \hat{q}_{2}, \hat{q}_{3}, \hat{p}_{1}, \hat{p}_{2}, \hat{p}_{3} \right)^{\mathrm{T}}, \\
        \left[ \hat{x}_{a}, \hat{x}_{b} \right] =&~ i\hbar \left( \bm{\Omega}_{3} \right)_{a,b},
    \end{aligned}
\end{align}
with $a,b\in\{1,\ldots,6\}$ and the symplectic form
\begin{align}
    \bm{\Omega}_{3} =&~
    \begin{pmatrix}
        \bm{0}_{3} & \bm{I}_{3} \\
        -\bm{I}_{3} & \bm{0}_{3}
    \end{pmatrix}.
\end{align}
The first moment vector and covariance matrix,
\begin{align}
    \begin{aligned}
        \bar{\bm{x}} =&~ \left\langle \hat{\bm{x}} \right\rangle, \\
        \left( \bm{\sigma} \right)_{a,b} =&~ \frac{1}{2} \left\langle \Delta\hat{x}_{a}\Delta\hat{x}_{b} + \Delta\hat{x}_{b}\Delta\hat{x}_{a} \right\rangle,
    \end{aligned}
\end{align}
with the fluctuation operator $\Delta\hat{x}_{a} = \hat{x}_{a} - \langle \hat{x}_{a} \rangle$, fully specify a Gaussian state.

In our framework, the pattern-resolved construction requires three ingredients: the entangled wire covariance, the readout covariance of the consumed nodes, and the Gaussian conditioning map to the retained mode.

We encode the three-node line graph, 
\begin{align}
    \bm{V}_{\mathrm{L}} =&~
    \begin{pmatrix}
        0 & 1 & 0 \\
        1 & 0 & 1 \\
        0 & 1 & 0
    \end{pmatrix},
\end{align}
through the controlled-phase network,
\begin{align}
    \hat{C}_{Z}(\bm{V}_{\mathrm{L}}) =&~ e^{ \frac{i}{2\hbar} \hat{\bm{q}}^{\mathrm{T}} \bm{V}_{\mathrm{L}} \hat{\bm{q}} },
\end{align}
whose symplectic form,
\begin{align}
    \bm{S}_{\mathrm{L}} =&~
    \begin{pmatrix}
        \bm{I}_{3} & \bm{0}_{3} \\
        \bm{V}_{\mathrm{L}} & \bm{I}_{3}
    \end{pmatrix},
\end{align}
acts on covariance matrices.
We inject an arbitrary one-mode Gaussian input,
\begin{align}
    \bm{\sigma}_{\mathrm{in}} =&~
    \begin{pmatrix}
        Q & C \\
        C & P
    \end{pmatrix},
    \qquad
    QP-C^{2}\geq\frac{\hbar^{2}}{4},
\end{align}
where $Q$ and $P$ are the input position and momentum variances, and $C$ is the position--momentum covariance.
We use independently momentum-squeezed resource modes at nodes $2$ and $3$, with
\begin{align}
    \begin{aligned}
        a = \frac{\hbar}{2}e^{2r}, \qquad
        \varepsilon = \frac{\hbar}{2}e^{-2r}, \qquad
         a\varepsilon = \frac{\hbar^{2}}{4}.
    \end{aligned}
\end{align}
The covariance before entangling,
\begin{align}
    \bm{\sigma}_{0} =&~
    \begin{pmatrix}
        Q & 0 & 0 & C & 0 & 0 \\
        0 & a & 0 & 0 & 0 & 0 \\
        0 & 0 & a & 0 & 0 & 0 \\
        C & 0 & 0 & P & 0 & 0 \\
        0 & 0 & 0 & 0 & \varepsilon & 0 \\
        0 & 0 & 0 & 0 & 0 & \varepsilon
    \end{pmatrix},
\end{align}
yields the entangled wire covariance,
\begin{align}
    \begin{aligned}
        \bm{\sigma}_{\mathrm{L}} =&~ \bm{S}_{\mathrm{L}} \bm{\sigma}_{0} \bm{S}_{\mathrm{L}}^{\mathrm{T}}  \\
        =&~
        \begin{pmatrix}
            Q & 0 & 0 & C & Q & 0 \\
            0 & a & 0 & a & 0 & a \\
            0 & 0 & a & 0 & a & 0 \\
            C & a & 0 & P+a & C & a \\
            Q & 0 & a & C & Q+a+\varepsilon & 0 \\
            0 & a & 0 & a & 0 & a+\varepsilon
        \end{pmatrix},
    \end{aligned}
\end{align}
which enters every pattern calculation.
The parameter $\varepsilon$ is the finite nullifier-noise variance that gives the squeezing-induced contribution to the output noise.

The readout covariance defines the second ingredient.
We use calibrated uncorrelated AK readout because it gives the minimal finite joint-quadrature Gaussian readout with a controlled momentum-homodyne limit~\cite{Appleby1998p1491,Arvind2020p126543}.
It contributes the Gaussian measurement covariance
\begin{align}
    \bm{\sigma}_{A} =&~ \operatorname{diag}\left( \nu_{q}, \nu_{p} \right),
\end{align}
where $\operatorname{diag}\left( \cdots \right)$ gives a diagonal matrix with the listed entries, and $\nu_{q}$ and $\nu_{p}$ are the effective variances of the joint position and momentum outcomes~\cite{Arthurs1965p725,Appleby1998p1491,Arvind2020p126543}.
The diagonal readout covariance isolates the uncorrelated AK calibration.
More general correlated AK readouts generate a broader family of Gaussian filters, but they are not needed to identify the operation-class change induced by finite joint-quadrature readout.
For pattern-resolved calculations, we allow the readout variances to depend on the consumed node and write them as $\nu_{qj}$ and $\nu_{pj}$ for $j=1,2$.
For a unit-gain minimum-noise AK calibration,
\begin{align}
    \nu_{q} = \frac{\hbar b}{2}, \qquad
    \nu_{p} = \frac{\hbar}{2b}, \qquad
    \nu_{q}\nu_{p} = \frac{\hbar^{2}}{4},
\end{align}
while $b=1$ gives
\begin{align}
    \nu_{q}=\nu_{p} =&~ \frac{\hbar}{2}.
\end{align}
We keep $\nu_{q}$ and $\nu_{p}$ explicit because these parameters decide whether the extra AK outcome remains part of the conditioning record or becomes uninformative.

For a consumed node $j=1,2$, the readout symbols,
\begin{align}
    \begin{aligned}
        \mathrm{H}_{j} &\equiv~ \hat{p}_{j}^{\prime}, \\
        \mathrm{A}_{j} &\equiv~ \left( \hat{q}_{j}^{\prime}, \hat{p}_{j}^{\prime} \right)^{\mathrm{T}},
    \end{aligned}
\end{align}
fix the measured vector $\hat{\bm{x}}_{M}^{(\mathcal{R})}$ and the measurement covariance $\bm{\sigma}_{M}^{(\mathcal{R})}$ for each pattern $\mathcal{R}$.
The retained variable,
\begin{align}
    \hat{\bm{x}}_{\mathrm{out}} =&~ \left( \hat{q}_{3}^{\prime}, \hat{p}_{3}^{\prime} \right)^{\mathrm{T}},
\end{align}
is common to all patterns.
We recover momentum homodyne readout in the singular Gaussian-measurement limit,
\begin{align}
    \nu_{q} \rightarrow \infty,
    \qquad
    \nu_{p} \rightarrow 0,
\end{align}
which keeps the momentum outcome sharp and removes information from the conjugate position outcome.
For a minimum-noise AK measurement, this limit is approached along
\begin{align}
    \nu_{q}\nu_{p} =&~ \frac{\hbar^{2}}{4}.
\end{align}
An AK node keeps both outcomes finite unless we explicitly coarse grain one component of the record.

The Gaussian conditioning rule provides the third ingredient~\cite{Weedbrook2012p621,Spedalieri2013p1350011,Genoni2016p331,Serafini2017book}.
For each pattern, the measured and retained variables define the block partition,
\begin{align}
    \bm{\sigma}_{\mathrm{L}}^{(\mathcal{R})} =&~
    \begin{pmatrix}
        \bm{A}_{\mathcal{R}} & \bm{C}_{\mathcal{R}} \\
        \bm{C}_{\mathcal{R}}^{\mathrm{T}} & \bm{B}_{\mathcal{R}}
    \end{pmatrix}.
\end{align}
The same partition gives the conditional first moments and covariance matrices,
\begin{align}
    \begin{aligned}
        \bar{\bm{x}}_{\mathrm{out}}^{(\mathcal{R})} =&~ \bar{\bm{x}}_{\mathrm{out}} + \bm{C}_{\mathcal{R}}^{\mathrm{T}} \left( \bm{A}_{\mathcal{R}} + \bm{\sigma}_{M}^{(\mathcal{R})} \right)^{-1} \left( \bm{m}_{\mathcal{R}} - \bar{\bm{x}}_{M}^{(\mathcal{R})} \right), \\
        \bm{\sigma}_{\mathrm{out}}^{(\mathcal{R})} =&~ \bm{B}_{\mathcal{R}} - \bm{C}_{\mathcal{R}}^{\mathrm{T}} \left( \bm{A}_{\mathcal{R}} + \bm{\sigma}_{M}^{(\mathcal{R})} \right)^{-1} \bm{C}_{\mathcal{R}},
    \end{aligned}
\end{align}
which define our pattern-resolved model.

\section{Pattern-resolved Gaussian maps}
\label{sec:Sec3}

We now evaluate the pattern-resolved model in the reduced finite-nullifier description of the cluster wire.
We first keep the anti-squeezed resource variance $a$ in the Schur-complement calculation and then take the reduced finite-nullifier description used for finitely squeezed cluster wires~\cite{Gu2009p062318,Alexander2014p062324,Ukai2011p240504}.
In this reduced description, all $a$-dependent terms cancel from the retained covariance transformation, while the finite nullifier variance $\varepsilon$ remains.
The homodyne pattern gives the reference additive Gaussian channel~\cite{Holevo2007p1,Weedbrook2012p621},
\begin{align}
    \begin{aligned}
        \bar{\bm{x}}_{\mathrm{out}}^{(\mathrm{HH})} =&~ \bm{G}_{\mathrm{HH}}\bar{\bm{x}}_{\mathrm{in}} + \bm{d}_{\mathrm{HH}}(\bm{m}_{\mathrm{HH}}), \\
        \bm{\sigma}_{\mathrm{out}}^{(\mathrm{HH})} =&~ \bm{G}_{\mathrm{HH}}\bm{\sigma}_{\mathrm{in}}\bm{G}_{\mathrm{HH}}^{\mathrm{T}} + \varepsilon\bm{I}_{2},
    \end{aligned}
\end{align}
with input-independent parity gain $\bm{G}_{\mathrm{HH}} = -\bm{I}_{2}$ and feed-forward displacement $\bm{d}_{\mathrm{HH}}(\bm{m}_{\mathrm{HH}}) = \left( m_{p2}, m_{p1} \right)^{\mathrm{T}}$.
Here $\bm{m}_{\mathrm{HH}} = \left( m_{p1}, m_{p2} \right)^{\mathrm{T}}$ is the homodyne outcome vector associated with $\hat{\bm{x}}_{M}^{(\mathrm{HH})} = \left( \hat{p}_{1}^{\prime}, \hat{p}_{2}^{\prime} \right)^{\mathrm{T}}$, and the ideal homodyne measurement covariance is $\bm{\sigma}_{M}^{(\mathrm{HH})} = \bm{0}_{2}$.
The homodyne pattern gives an input-independent additive Gaussian channel with parity gain $\bm{G}_{\mathrm{HH}}$ and finite-squeezing noise $\varepsilon\bm{I}_{2}$.

The mixed patterns differ from the reference parity wire because a finite AK readout keeps the extra conjugate outcome in the conditioning record.
We only need to calculate the fully resolved $\mathrm{AA}$ pattern with node-resolved readout noises, since the $\mathrm{AH}$, $\mathrm{HA}$, and $\mathrm{HH}$ patterns follow by taking the corresponding homodyne limits.

For the $\mathrm{AA}$ pattern, both consumed nodes carry AK readout.
The measured vector, outcome vector, and measurement covariance,
\begin{align}
    \begin{aligned}
        \hat{\bm{x}}_{M}^{(\mathrm{AA})} =&~ \left( \hat{q}_{1}^{\prime}, \hat{q}_{2}^{\prime}, \hat{p}_{1}^{\prime}, \hat{p}_{2}^{\prime} \right)^{\mathrm{T}}, \\
        \bm{m}_{\mathrm{AA}} =&~ \left( m_{q1}, m_{q2}, m_{p1}, m_{p2} \right)^{\mathrm{T}}, \\
        \bm{\sigma}_{M}^{(\mathrm{AA})} =&~ \operatorname{diag}\left( \nu_{q1},  \nu_{q2}, \nu_{p1}, \nu_{p2} \right).
    \end{aligned}
\end{align}
These objects yield the conditional first moments and covariance,
\begin{align}
    \begin{aligned}
        \bar{\bm{x}}_{\mathrm{out}}^{(\mathrm{AA})} =&~ \bm{G}_{\mathrm{AA}}\bar{\bm{x}}_{\mathrm{in}} + \bm{d}_{\mathrm{AA}}(\bm{m}_{\mathrm{AA}}), \\
        \bm{\sigma}_{\mathrm{out}}^{(\mathrm{AA})} =&~
        \bm{G}_{\mathrm{AA}}\bm{\sigma}_{\mathrm{in}}\bm{G}_{\mathrm{AA}}^{\mathrm{T}} + \bm{N}_{\mathrm{AA}}^{(\mathrm{cond})},
    \end{aligned}
\end{align}
with retained covariance, finite-readout gain, feed-forward displacement, and residual conditional covariance,
\begin{align}
   \begin{aligned}
        \bm{\sigma}_{\mathrm{out}}^{(\mathrm{AA})} =&~ \frac{1}{\Delta_{\mathrm{AA}}}
        \begin{pmatrix}
            \sigma_{qq}^{(\mathrm{AA})} & \sigma_{qp}^{(\mathrm{AA})} \\
            \sigma_{qp}^{(\mathrm{AA})} & \sigma_{pp}^{(\mathrm{AA})}
        \end{pmatrix}, \\
        \bm{G}_{\mathrm{AA}} =&~
        \begin{pmatrix}
            \frac{D_{1}}{\Delta_{\mathrm{AA}}}-1 &
            \frac{C\nu_{q1}}{\Delta_{\mathrm{AA}}} \\
            \frac{C\nu_{q2}}{\Delta_{\mathrm{AA}}} &
            \frac{D_{2}}{\Delta_{\mathrm{AA}}}-1
        \end{pmatrix}, \\
        \bm{d}_{\mathrm{AA}}(\bm{m}_{\mathrm{AA}}) =&~
        \begin{pmatrix}
            - \frac{D_{1}}{\Delta_{\mathrm{AA}}}m_{q1}
            + \frac{C\nu_{q1}}{\Delta_{\mathrm{AA}}}m_{q2}
            - \frac{C\nu_{q1}}{\Delta_{\mathrm{AA}}}m_{p1}
            + m_{p2}
            \\
            - \frac{C\nu_{q2}}{\Delta_{\mathrm{AA}}}m_{q1}
            + \frac{D_{2}}{\Delta_{\mathrm{AA}}}m_{q2}
            + \left( 1-\frac{D_{2}}{\Delta_{\mathrm{AA}}} \right)m_{p1}
        \end{pmatrix}, \\
        \bm{N}_{\mathrm{AA}}^{(\mathrm{cond})} =&~ \frac{1}{\Delta_{\mathrm{AA}}^{2}}
        \begin{pmatrix}
            N_{qq}^{(\mathrm{AA})} & N_{qp}^{(\mathrm{AA})} \\
            N_{qp}^{(\mathrm{AA})} & N_{pp}^{(\mathrm{AA})}
        \end{pmatrix},
    \end{aligned}
\end{align}
the covariance matrix elements, 
\begin{align}
    \begin{aligned}
        \sigma_{qq}^{(\mathrm{AA})} =&~ \left( \varepsilon+\nu_{p2} \right)\Delta_{\mathrm{AA}}+\nu_{q1}D_{1}, \\
        \sigma_{qp}^{(\mathrm{AA})} =&~ C\nu_{q1}\nu_{q2}, \\
        \sigma_{pp}^{(\mathrm{AA})} =&~ \varepsilon\Delta_{\mathrm{AA}}+\nu_{q2}D_{2},
    \end{aligned}
\end{align}
the feed-forward factors and Schur denominator,
\begin{align}
    \begin{aligned}
        D_{1} =&~ \Delta_{\mathrm{AA}}-\nu_{q1}\left( P+\nu_{p1}+\nu_{q2} \right), \\
        D_{2} =&~ \Delta_{\mathrm{AA}}-\nu_{q2}\left( Q+\nu_{q1} \right), \\
        \Delta_{\mathrm{AA}} =&~ \left( Q+\nu_{q1} \right)\left( P+\nu_{p1}+\nu_{q2} \right)-C^{2},
    \end{aligned}
\end{align}
and the residual entries
\begin{align}
    \begin{aligned}
        N_{qq}^{(\mathrm{AA})} =&~
        \left( \varepsilon+\nu_{p2}+\nu_{q1} \right)\Delta_{\mathrm{AA}}^{2} \\
        &+ C^{2}
        \left[
            2\nu_{q1}\left( \Delta_{\mathrm{AA}}-D_{1} \right)
            +\nu_{q1}^{2}\left( \nu_{p1}+\nu_{q2} \right)
        \right] \\
        &-\left( \Delta_{\mathrm{AA}}-D_{1} \right)^{2}
        \left[
            \frac{2\left( \Delta_{\mathrm{AA}}-D_{2} \right)}{\nu_{q2}}
            -\nu_{q1}
        \right], \\
        N_{qp}^{(\mathrm{AA})} =&~
        C\nu_{q1}
        \left[
            \nu_{q2}\left( \Delta_{\mathrm{AA}}+D_{1} \right)
            - \right. \\
            &~ \left. \left( \nu_{p1}+\nu_{q2} \right)
            \left( \Delta_{\mathrm{AA}}-D_{2} \right)
        \right], \\
        N_{pp}^{(\mathrm{AA})} =&~
        \left( \varepsilon+\nu_{q2} \right)\Delta_{\mathrm{AA}}^{2} \\
        &+ C^{2}
        \left[
            2\nu_{q2}\left( \Delta_{\mathrm{AA}}-D_{2} \right)
            +\nu_{q1}\nu_{q2}^{2}
        \right] \\
        &-\left( \Delta_{\mathrm{AA}}-D_{2} \right)^{2}
        \left[
            \frac{2\left( \Delta_{\mathrm{AA}}-D_{1} \right)}{\nu_{q1}}
            -\nu_{p1}-\nu_{q2}
        \right].
    \end{aligned}
\end{align}
The $\mathrm{AA}$ displacement shows how the two finite AK position records enter the feed-forward correction asymmetrically.
The node-$1$ position record $m_{q1}$ enters both output quadratures through $D_{1}$ and $C\nu_{q2}$, while the node-$2$ position record $m_{q2}$ enters through $C\nu_{q1}$ and $D_{2}$.
The singular homodyne limits remove the corresponding position record and reduce these terms to the mixed-pattern displacements.

The homodyne limit on node $j\in\left\{ 1,2 \right\}$ removes the AK position outcome from the conditioning record and keeps the momentum outcome sharp.
For minimum-noise readout, the limit $\mathcal{H}_{j}$ uses the singular path $\nu_{qj}\rightarrow\infty$ and $\nu_{pj}\rightarrow0$, with
\begin{align}
    \nu_{qj}\nu_{pj} =&~ \frac{\hbar^{2}}{4}.
\end{align}
Direct singular-path substitution in the node-resolved $\mathrm{AA}$ expressions gives the remaining patterns.
The limits $\mathcal{H}_{2}$, $\mathcal{H}_{1}$, and $\mathcal{H}_{1}\mathcal{H}_{2}$ give $\mathrm{AH}$, $\mathrm{HA}$, and $\mathrm{HH}$, respectively, recovering the reference additive Gaussian channel with parity gain $\bm{G}_{\mathrm{HH}}=-\bm{I}_{2}$ and noise $\varepsilon\bm{I}_{2}$.

For the $\mathrm{AH}$ pattern, applying $\mathcal{H}_{2}$ deletes the uninformative node-$2$ position record,
\begin{align}
    \begin{aligned}
        \hat{\bm{x}}_{M}^{(\mathrm{AH})} =&~ \left( \hat{q}_{1}^{\prime}, \hat{p}_{1}^{\prime}, \hat{p}_{2}^{\prime} \right)^{\mathrm{T}}, \\
        \bm{m}_{\mathrm{AH}} =&~ \left( m_{q1}, m_{p1}, m_{p2} \right)^{\mathrm{T}}, \\
        \bm{\sigma}_{M}^{(\mathrm{AH})} =&~ \operatorname{diag}\left( \nu_{q1}, \nu_{p1}, 0 \right),
    \end{aligned}
\end{align}
has the conditional first moments and covariance
\begin{align}
    \begin{aligned}
        \bar{\bm{x}}_{\mathrm{out}}^{(\mathrm{AH})} =&~ \bm{G}_{\mathrm{AH}}\bar{\bm{x}}_{\mathrm{in}} + \bm{d}_{\mathrm{AH}}(\bm{m}_{\mathrm{AH}}), \\
        \bm{\sigma}_{\mathrm{out}}^{(\mathrm{AH})} =&~
        \bm{G}_{\mathrm{AH}}\bm{\sigma}_{\mathrm{in}}\bm{G}_{\mathrm{AH}}^{\mathrm{T}}
        +
        \bm{N}_{\mathrm{AH}}^{(\mathrm{cond})},
    \end{aligned}
\end{align}
with conditional covariance matrix, finite-readout gain, feed-forward displacement, and residual conditional covariance,
\begin{align}
    \begin{aligned}
        \bm{\sigma}_{\mathrm{out}}^{(\mathrm{AH})} =&~
        \frac{1}{Q+\nu_{q1}}
        \begin{pmatrix}
            \sigma_{qq}^{(\mathrm{AH})} & \sigma_{qp}^{(\mathrm{AH})} \\
            \sigma_{qp}^{(\mathrm{AH})} & \sigma_{pp}^{(\mathrm{AH})}
        \end{pmatrix}, \\
        \bm{G}_{\mathrm{AH}} =&~
        \begin{pmatrix}
            -\frac{\nu_{q1}}{Q+\nu_{q1}} & 0 \\
            \frac{C}{Q+\nu_{q1}} & -1
        \end{pmatrix}, \\
        \bm{d}_{\mathrm{AH}}(\bm{m}_{\mathrm{AH}}) =&~
        \begin{pmatrix}
            m_{p2}-\frac{Q}{Q+\nu_{q1}}m_{q1} \\
            m_{p1}-\frac{C}{Q+\nu_{q1}}m_{q1}
        \end{pmatrix}, \\
        \bm{N}_{\mathrm{AH}}^{(\mathrm{cond})} =&~
        \frac{1}{\left( Q+\nu_{q1} \right)^{2}}
        \begin{pmatrix}
            N_{qq}^{(\mathrm{AH})} & N_{qp}^{(\mathrm{AH})} \\
            N_{qp}^{(\mathrm{AH})} & N_{pp}^{(\mathrm{AH})}
        \end{pmatrix},
    \end{aligned}
\end{align}
the covariance matrix elements,
\begin{align}
    \begin{aligned}
        \sigma_{qq}^{(\mathrm{AH})} =&~ \varepsilon\left( Q+\nu_{q1} \right)+\nu_{q1}Q, \\
        \sigma_{qp}^{(\mathrm{AH})} =&~ C\nu_{q1}, \\
        \sigma_{pp}^{(\mathrm{AH})} =&~ \left( \varepsilon+\nu_{p1}+P \right)\left( Q+\nu_{q1} \right)-C^{2},
    \end{aligned}
\end{align}
and the residual entries
\begin{align}
    \begin{aligned}
        N_{qq}^{(\mathrm{AH})} =&~
        \varepsilon\left( Q+\nu_{q1} \right)^{2}
        +\nu_{q1}Q^{2}, \\
        N_{qp}^{(\mathrm{AH})} =&~
        C\nu_{q1}Q, \\
        N_{pp}^{(\mathrm{AH})} =&~
        \left( \varepsilon+\nu_{p1} \right)\left( Q+\nu_{q1} \right)^{2}
        +C^{2}\nu_{q1}.
    \end{aligned}
\end{align}
The $\mathrm{AH}$ pattern keeps the conditional filter generated by the finite AK position readout on node $1$.
Its gain, conditional covariance, and residual covariance depend on the input position variance $Q$ and on the input position--momentum covariance $C$.

For the $\mathrm{HA}$ pattern, applying $\mathcal{H}_{1}$ deletes the uninformative node-$1$ position record,
\begin{align}
    \begin{aligned}
        \hat{\bm{x}}_{M}^{(\mathrm{HA})} =&~ \left( \hat{q}_{2}^{\prime}, \hat{p}_{1}^{\prime},  \hat{p}_{2}^{\prime} \right)^{\mathrm{T}}, \\
        \bm{m}_{\mathrm{HA}} =&~ \left( m_{q2}, m_{p1},  m_{p2} \right)^{\mathrm{T}}, \\
        \bm{\sigma}_{M}^{(\mathrm{HA})} =&~ \operatorname{diag}\left(\nu_{q2}, 0,  \nu_{p2} \right),
    \end{aligned}
\end{align}
has the conditional first moments and covariance
\begin{align}
    \begin{aligned}
        \bar{\bm{x}}_{\mathrm{out}}^{(\mathrm{HA})} =&~ \bm{G}_{\mathrm{HA}}\bar{\bm{x}}_{\mathrm{in}} + \bm{d}_{\mathrm{HA}}(\bm{m}_{\mathrm{HA}}), \\
        \bm{\sigma}_{\mathrm{out}}^{(\mathrm{HA})} =&~
        \bm{G}_{\mathrm{HA}}\bm{\sigma}_{\mathrm{in}}\bm{G}_{\mathrm{HA}}^{\mathrm{T}}
        +
        \bm{N}_{\mathrm{HA}}^{(\mathrm{cond})},
    \end{aligned}
\end{align}
with conditional covariance matrix, finite-readout gain, feed-forward displacement, and residual conditional covariance,
\begin{align}
    \begin{aligned}
        \bm{\sigma}_{\mathrm{out}}^{(\mathrm{HA})} =&~
        \frac{1}{P+\nu_{q2}}
        \begin{pmatrix}
            \sigma_{qq}^{(\mathrm{HA})} & \sigma_{qp}^{(\mathrm{HA})} \\
            \sigma_{qp}^{(\mathrm{HA})} & \sigma_{pp}^{(\mathrm{HA})}
        \end{pmatrix}, \\
        \bm{G}_{\mathrm{HA}} =&~
        \begin{pmatrix}
            -1 & \frac{C}{P+\nu_{q2}} \\
            0 & -\frac{\nu_{q2}}{P+\nu_{q2}}
        \end{pmatrix}, \\
        \bm{d}_{\mathrm{HA}}(\bm{m}_{\mathrm{HA}}) =&~
        \begin{pmatrix}
            m_{p2}-\frac{C}{P+\nu_{q2}}m_{p1}+\frac{C}{P+\nu_{q2}}m_{q2} \\
            \frac{\nu_{q2}}{P+\nu_{q2}}m_{p1}+\frac{P}{P+\nu_{q2}}m_{q2}
        \end{pmatrix}, \\
        \bm{N}_{\mathrm{HA}}^{(\mathrm{cond})} =&~
        \frac{1}{\left( P+\nu_{q2} \right)^{2}}
        \begin{pmatrix}
            N_{qq}^{(\mathrm{HA})} & N_{qp}^{(\mathrm{HA})} \\
            N_{qp}^{(\mathrm{HA})} & N_{pp}^{(\mathrm{HA})}
        \end{pmatrix},
    \end{aligned}
\end{align}
the covariance matrix elements,
\begin{align}
    \begin{aligned}
        \sigma_{qq}^{(\mathrm{HA})} =&~ \left( \varepsilon+\nu_{p2}+Q \right)\left( P+\nu_{q2} \right)-C^{2}, \\
        \sigma_{qp}^{(\mathrm{HA})} =&~ C\nu_{q2}, \\
        \sigma_{pp}^{(\mathrm{HA})} =&~ \varepsilon\left( P+\nu_{q2} \right)+\nu_{q2}P,
    \end{aligned}
\end{align}
and the residual entries
\begin{align}
    \begin{aligned}
        N_{qq}^{(\mathrm{HA})} =&~
        \left( \varepsilon+\nu_{p2} \right)\left( P+\nu_{q2} \right)^{2}
        +C^{2}\nu_{q2}, \\
        N_{qp}^{(\mathrm{HA})} =&~
        CP\nu_{q2}, \\
        N_{pp}^{(\mathrm{HA})} =&~
        \varepsilon\left( P+\nu_{q2} \right)^{2}
        +P^{2}\nu_{q2}.
    \end{aligned}
\end{align}
The $\mathrm{HA}$ pattern gives the complementary conditional filter.
Its finite AK position readout acts through the node-$2$ conditioning record, so the gain, conditional covariance, and residual covariance depend on the input momentum variance $P$ and on the input position--momentum covariance $C$.

Our pattern-resolved maps separate the input-independent homodyne reference from the finite-readout AK patterns.
The homodyne pattern has a fixed gain and a fixed additive-noise matrix.
By contrast, any finite AK position readout keeps an extra conditioning variable and changes the Schur complement.
For $\mathcal{R}\in\left\{ \mathrm{AA}, \mathrm{AH}, \mathrm{HA} \right\}$, the finite-readout transformations can be written in the conditional form
\begin{align}
    \begin{aligned}
        \bar{\bm{x}}_{\mathrm{out}}^{(\mathcal{R})} =&~
        \bm{G}_{\mathcal{R}}\left( \bm{\sigma}_{\mathrm{in}} \right)
        \bar{\bm{x}}_{\mathrm{in}}
        +
        \bm{d}_{\mathcal{R}}\left( \bm{m}_{\mathcal{R}};\bm{\sigma}_{\mathrm{in}} \right), \\
        \bm{\sigma}_{\mathrm{out}}^{(\mathcal{R})} =&~
        \bm{G}_{\mathcal{R}}\left( \bm{\sigma}_{\mathrm{in}} \right)
        \bm{\sigma}_{\mathrm{in}}
        \bm{G}_{\mathcal{R}}^{\mathrm{T}}\left( \bm{\sigma}_{\mathrm{in}} \right)
        +
        \bm{N}_{\mathcal{R}}^{(\mathrm{cond})}\left( \bm{\sigma}_{\mathrm{in}} \right),
    \end{aligned}
\end{align}
with explicit dependence on the independent covariance entries $Q$, $P$, and $C$,
\begin{align}
    \frac{\partial \bm{G}_{\mathcal{R}}}{\partial \bm{\sigma}_{\mathrm{in}}}\neq\bm{0},
    \qquad
    \frac{\partial \bm{N}_{\mathcal{R}}^{(\mathrm{cond})}}{\partial \bm{\sigma}_{\mathrm{in}}}\neq\bm{0}.
\end{align}
No input-independent pair $\left( \bm{G}_{\mathcal{R}},\bm{N}_{\mathcal{R}} \right)$ defines the finite-AK patterns as additive Gaussian channels.
The readout pattern instead selects the covariance sector entering the conditional filter.

\section{Conditional-filter diagnostics}
\label{sec:Sec4}

We now diagnose the conditional Gaussian filters produced by the pattern-resolved maps.
The homodyne pattern gives the additive reference channel.
The finite-AK patterns instead retain one or two extra position outcomes in the conditioning record, so the Schur complement depends on the input covariance.
This distinction separates additive teleportation noise from covariance-selective Gaussian filtering.

For a one-mode covariance matrix $\bm{\sigma}$, the Williamson scalar~\cite{Simon1994p1567,Weedbrook2012p621,Serafini2017book},
\begin{align}
    \nu_{W}\left( \bm{\sigma} \right) =&~ \sqrt{\det \bm{\sigma}},
\end{align}
measures the symplectic area of the covariance ellipse and obeys
\begin{align}
    \nu_{W}\left( \bm{\sigma} \right) \geq \frac{\hbar}{2}.
\end{align}
We use this scalar to compare the covariance area of the input state, 
\begin{align}
    \nu_{\mathrm{in}} =&~ \left( QP-C^{2} \right)^{1/2}
    \geq \frac{\hbar}{2},
\end{align}
with that of the retained output state for each readout pattern,
\begin{align}
    \begin{aligned}
        \nu_{\mathrm{out}}^{(\mathrm{HH})} =&~ \left[ \left( Q+\varepsilon \right) \left( P + \varepsilon \right) - C^{2} \right]^{1/2}, \\
        \nu_{\mathrm{out}}^{(\mathrm{AA})} =&~ \frac{1}{\Delta_{\mathrm{AA}}} \left[ \sigma_{qq}^{(\mathrm{AA})} \sigma_{pp}^{(\mathrm{AA})} - \left( \sigma_{qp}^{(\mathrm{AA})} \right)^{2} \right]^{1/2}, \\
        \nu_{\mathrm{out}}^{(\mathrm{AH})} =&~ \left[ \frac{ \left( \varepsilon + \nu_{p1} + P \right) \left[ \varepsilon \left( Q + \nu_{q1} \right) + \nu_{q1} Q \right] - C^{2} \left( \varepsilon + \nu_{q1} \right) } { Q + \nu_{q1} } \right]^{1/2}, \\
        \nu_{\mathrm{out}}^{(\mathrm{HA})} =&~ \left[ \frac{ \left( \varepsilon + \nu_{p2} + Q \right) \left[ \varepsilon\left( P + \nu_{q2} \right) + \nu_{q2} P \right] - C^{2} \left( \varepsilon + \nu_{q2} \right) } { P + \nu_{q2} } \right]^{1/2},
    \end{aligned}
\end{align}
through the output-to-input symplectic-area ratio,
\begin{align}
    \eta_{\mathcal{R}} =&~
    \frac{\nu_{\mathrm{out}}^{(\mathcal{R})}}{\nu_{\mathrm{in}}},
\end{align}
which measures the direct area deformation of the output state with respect to the input state.
Values $\eta_{\mathcal{R}}>1$ indicate output-area expansion, while values $\eta_{\mathcal{R}}<1$ indicate conditional output-area reduction,
\begin{align}
    \begin{aligned}
        \eta_{\mathrm{HH}} =&~ \left[ \frac{ \left( Q + \varepsilon \right) \left( P + \varepsilon \right) - C^{2} }{ QP-C^{2} } \right]^{1/2}, \\
        \eta_{\mathrm{AA}} =&~ \frac{1}{\Delta_{\mathrm{AA}} \left( Q P - C^{2} \right)^{1/2} } \left[ \sigma_{qq}^{(\mathrm{AA})} \sigma_{pp}^{(\mathrm{AA})} - \left( \sigma_{qp}^{(\mathrm{AA})} \right)^{2} \right]^{1/2}, \\
        \eta_{\mathrm{AH}} =&~ \left[ \frac{ \left( \varepsilon + \nu_{p1} + P \right) \left[ \varepsilon \left( Q + \nu_{q1} \right) + \nu_{q1} Q \right] - C^{2} \left( \varepsilon + \nu_{q1} \right) }{ \left( Q + \nu_{q1} \right) \left( Q P - C^{2} \right) } \right]^{1/2}, \\
        \eta_{\mathrm{HA}} =&~ \left[ \frac{ \left( \varepsilon + \nu_{p2} + Q \right) \left[ \varepsilon \left( P + \nu_{q2} \right) + \nu_{q2} P \right] - C^{2} \left( \varepsilon + \nu_{q2} \right) } { \left( P+\nu_{q2} \right) \left( Q P - C^{2} \right) } \right]^{1/2}.
    \end{aligned}
\end{align}
The homodyne pattern gives the additive-noise area expansion of the reference teleportation channel and reduces to unity in the absence of finite-squeezing noise.
The finite-AK patterns instead give covariance-dependent area deformations generated by the Schur-complement filter.

\begin{figure}[t]
    \centering
    \includegraphics[width= 0.75 \textwidth]{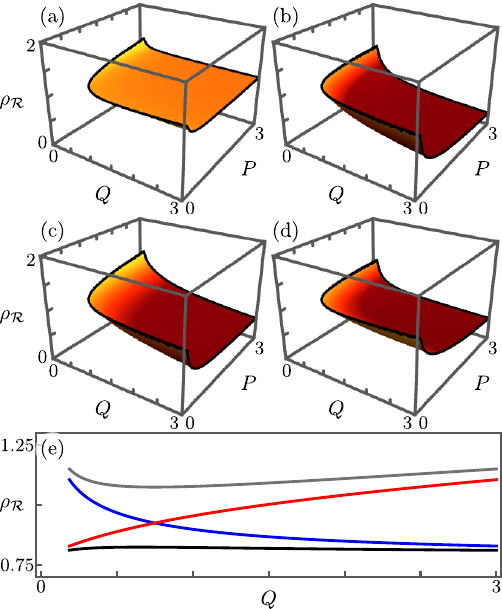}
    \caption{
        Output-to-input symplectic-area ratio $\eta_{\mathcal{R}}$ for readout patterns (a) $\mathrm{HH}$, (b) $\mathrm{AA}$, (c) $\mathrm{AH}$, and (d) $\mathrm{HA}$ over the physical input-covariance region $Q P - C^{2} \geq \hbar^{2}/4$.
        (e) Line cut along $P(Q) = \left( C^{2} + \hbar^{2}/2 \right)/Q$, so $Q P - C^{2} = \hbar^{2}/2$ remains fixed.
        The gray, black, blue, and red curves correspond to $\mathrm{HH}$, $\mathrm{AA}$, $\mathrm{AH}$, and $\mathrm{HA}$.
    }
    \label{fig:Fig2}
\end{figure}

Figure~\ref{fig:Fig2} shows $\eta_{\mathcal{R}}$ over the physical input-covariance domain $Q P - C^{2} \geq \hbar^{2}/ 4$.
We fix $\varepsilon = 0.05~\hbar$, use symmetric node-resolved AK readout variances $\nu_{q1}=\nu_{p1}=\nu_{q2}=\nu_{p2} = \hbar/2$, and set $C = \hbar / 4 $ while scanning $Q, P \in \left[ 0, 3 \right]~\hbar$.
Over this domain, the numerical ranges are $\eta_{\mathrm{HH}} \in \left[ 1.0168, 1.2770 \right]$, $\eta_{\mathrm{AA}} \in \left[ 0.2309, 1.1177 \right]$, $\eta_{\mathrm{AH}} \in \left[ 0.4347, 1.3063 \right]$, and $\eta_{\mathrm{HA}} \in \left[ 0.4347, 1.3063 \right]$.
The homodyne pattern expands the input symplectic area because finite squeezing adds noise to the parity channel.
The finite-AK patterns can either expand or reduce the retained-output area, depending on the input covariance.
Finite AK readout does not act as additive noise on the retained state.
It implements a conditional Gaussian filter whose output area depends on the covariance sector selected by the measurement record.

We compare the patterns along a fixed input-area trajectory $P(Q) = \left( C^{2} + \hbar^{2}/2 \right) / Q$ in Fig.~\ref{fig:Fig2}(e).
This trajectory keeps $Q P - C^{2} = \frac{\hbar^{2}}{2}$ fixed and scans $Q \in \left[ 0.1875, 3 \right]~\hbar$, while $P$ decreases from $3~\hbar$ to $0.1875~\hbar$.
Along this cut, the numerical ranges are $\eta_{\mathrm{HH}} \in \left[ 1.0747, 1.1505 \right]$, $\eta_{\mathrm{AA}} \in \left[ 0.8105, 0.8239 \right]$, $\eta_{\mathrm{AH}} \in \left[ 0.8284, 1.1059 \right]$, and $\eta_{\mathrm{HA}} \in \left[ 0.8284, 1.1059 \right]$.
The $\mathrm{AA}$ pattern remains below unity along this cut, so the fully resolved AK record reduces the retained symplectic area relative to the input.
The mixed patterns cross from area reduction to area expansion as the cut exchanges position and momentum variance.
This behavior reflects the analytic structure of the filters, where $\mathrm{AH}$ selects the position sector through $Q$ and $C$, $\mathrm{HA}$ selects the momentum sector through $P$ and $C$, and $\mathrm{AA}$ retains both covariance sectors through $\Delta_{\mathrm{AA}}$.

To separate the net deformation measured by $\eta_{\mathcal{R}}$ from the area transferred by the conditional linear gain, we track the gain determinant~\cite{Holevo2007p1,Caruso2006p310,Weedbrook2012p621},
\begin{align}
    \tau_{\mathcal{R}} =&~ \det\bm{G}_{\mathcal{R}}.
\end{align}
For our readout patterns, $\tau_{\mathcal{R}}\geq0$, so this determinant gives the Williamson-area factor of the gain-propagated covariance contribution,
\begin{align}
    \det\left( \bm{G}_{\mathcal{R}}\bm{\sigma}_{\mathrm{in}}\bm{G}_{\mathcal{R}}^{\mathrm{T}} \right)
    =&~
    \tau_{\mathcal{R}}^{2}\det\bm{\sigma}_{\mathrm{in}}.
\end{align}
The four patterns give
\begin{align}
    \begin{aligned}
        \tau_{\mathrm{HH}} =&~ 1, \\
        \tau_{\mathrm{AA}} =&~ \frac{\nu_{q1}\nu_{q2}}{\Delta_{\mathrm{AA}}}, \\
        \tau_{\mathrm{AH}} =&~ \frac{\nu_{q1}}{Q+\nu_{q1}}, \\
        \tau_{\mathrm{HA}} =&~ \frac{\nu_{q2}}{P+\nu_{q2}}.
    \end{aligned}
\end{align}
For positive finite readout variances and physical input covariances satisfying $QP-C^{2} \geq \hbar^{2}/4$, the finite-AK patterns satisfy $\tau_{\mathcal{R}}\in\left( 0,1 \right)$ and obey
\begin{align}
    0<\tau_{\mathrm{AA}}<\min\left\{ \tau_{\mathrm{AH}},\tau_{\mathrm{HA}} \right\}<1.
\end{align}

\begin{figure}[t]
    \centering
    \includegraphics[width= 0.75 \textwidth]{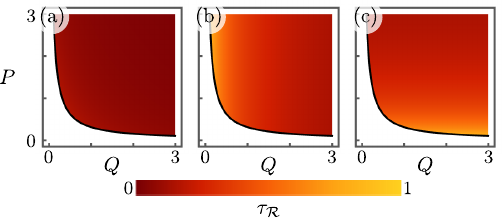}
    \caption{
        Gain determinant $\tau_{\mathcal{R}}$ for readout patterns (a) $\mathrm{AA}$, (b) $\mathrm{AH}$, and (c) $\mathrm{HA}$ over the physical input-covariance region $Q P - C^{2} \geq \hbar^{2} / 4$, with the black curve marking the boundary.
        We use the fixed color scale $\tau_{\mathcal{R}} \in \left[ 0, 1 \right]$ and omit the constant homodyne pattern $\tau_{\mathrm{HH}} = 1$.
    }
    \label{fig:Fig3}
\end{figure}

Figure~\ref{fig:Fig3} shows $\tau_{\mathcal{R}}$ over the same physical domain and with the same parameters as Fig.~\ref{fig:Fig2}.
All plots use the fixed color scale $\tau_{\mathcal{R}} \in \left[ 0, 1 \right]$.
Over this domain, the numerical ranges are $\tau_{\mathrm{AA}} \in \left[ 0.0180, 0.1622 \right]$, $\tau_{\mathrm{AH}} \in \left[ 0.1428, 0.8276 \right]$, and $\tau_{\mathrm{HA}} \in \left[ 0.1428, 0.8276 \right]$.
The $\mathrm{AH}$ and $\mathrm{HA}$ ranges coincide because the scan domain and the symmetric readout calibration are invariant under $Q\leftrightarrow P$, while $\tau_{\mathrm{AH}}$ and $\tau_{\mathrm{HA}}$ transform into one another under this exchange.
The $\mathrm{AA}$ pattern has the strongest gain-area contraction because both finite position records enter the two-node Schur denominator $\Delta_{\mathrm{AA}}$.
The mixed patterns retain larger gain determinants because only one finite position record enters the conditioning step.

The output-to-input symplectic-area ratio $\eta_{\mathcal{R}}$ and the gain determinant $\tau_{\mathcal{R}}$ separate the net output-area deformation from the area transferred by the conditional gain.
The homodyne pattern changes the output area only through additive finite-squeezing noise.
The finite-AK patterns instead combine gain-area contraction with covariance-dependent residual conditioning.
An output-area reduction $\eta_{\mathcal{R}}<1$ should not be interpreted as reduced additive noise.
It signals a conditional filtering effect generated by the retained finite-AK measurement record.

\section{Conclusion}
\label{sec:Sec5}

We classified the Gaussian transformations generated by calibrated uncorrelated Arthurs--Kelly readout in a three-node continuous-variable cluster wire.
The homodyne pattern $\mathrm{HH}$ gives the reference additive Gaussian channel, implementing the parity channel with fixed parity gain and additive finite-squeezing noise.
Replacing one or both consumed homodyne readouts by finite AK readouts changes the operation class.

The finite-AK readout patterns retain one or two extra position outcomes in the conditioning record.
These outcomes enter the Schur complement and make the retained covariance depend on the input covariance entries $Q$, $P$, and $C$.
The $\mathrm{AA}$, $\mathrm{AH}$, and $\mathrm{HA}$ readout patterns preserve Gaussianity but do not define input-independent additive Gaussian channels.
They implement covariance-dependent conditional Gaussian filters.

The readout pattern selects the filtered covariance sector.
The $\mathrm{AH}$ pattern acts through the position sector, the $\mathrm{HA}$ pattern acts through the momentum sector, and the fully resolved $\mathrm{AA}$ pattern retains both sectors.
The singular AK homodyne path recovers the homodyne pattern when the position record becomes uninformative and the momentum outcome becomes sharp.

The gain determinant gives a compact quantitative signature of the finite-readout filters.
For positive finite readout variances and physical input covariances satisfying $QP-C^{2}\geq  \hbar^{2} / 4$, the finite-AK patterns obey
\begin{align}
    0<\tau_{\mathrm{AA}}<\min\left\{ \tau_{\mathrm{AH}},\tau_{\mathrm{HA}} \right\}<1.
\end{align}
The fully resolved AK pattern has the strongest gain-area contraction because both finite position records enter the two-node Schur denominator.

The output-to-input area ratio and gain-area diagnostics show the structure from complementary viewpoints.
Finite-AK readout combines gain contraction with covariance-dependent residual conditioning, rather than adding a fixed noise matrix to the retained mode.
Our results identify calibrated uncorrelated AK readout as a measurement-level method for programming covariance-sensitive Gaussian filters inside a fixed cluster graph.
The graph and resource covariance remain unchanged, while the readout pattern and calibration determine which covariance entries enter the filter.


\section*{Funding}
Not Applicable.

\section*{Acknowledgments}
B.~M.~R.-L. thanks Jacinta Alderete Galan for daycare support during this work and acknowledges support and hospitality as an affiliate visiting colleague at the Department of Physics and Astronomy, University of New Mexico.

\section*{Disclosures}
The authors declare no conflicts of interest.

\section*{Data Availability Statement}
All data generated or analyzed during this study are available from the corresponding author upon reasonable request.



%

\end{document}